\documentclass[11pt]{article}

\def\Msun{\hbox{$M_\odot$}}

\def\m{\multicolumn}

\usepackage{makeidx}  
\usepackage{graphicx} 
\usepackage{multicol} 

\begin{document}

\large
\begin{center}
{\bf A study of the old  galactic star cluster Berkeley 32 }

\vspace{0.3cm}

\normalsize
 {\sf T. Richtler$^{1}$ and R. Sagar$^{2,3}$} \\
{\it$1$ Departamento de F\'{\i}sica, Universidad de Concepci\'on, Casilla 160-C,  Concepci\'on, Chile,  \\ e-mail: tom@coma.cfm.udec.cl} \\ 
{\it$2$ U. P. State Observatory, Manora Peak, Naini Tal - 263 129, India} \\
{\it$3$ Sternwarte der Universit\"{a}t Bonn, Auf dem H\"{u}gel 71, D-53121 Bonn, Germany} \\
\end{center}

\begin{abstract}
We present new CCD photometry of the distant old open star cluster
Berkeley 32 in Johnson $V$ and Cousins $I$ passbands. A total of $\sim$ 3200 
stars have been observed in a field of about $13^{'} \times 13^{'}$. 
The colour-magnitude diagram (CMD) in $V, (V-I)$ has been generated down 
to V = 22 mag. A broad but well defined main sequence is clearly visible. Some 
blue stragglers, a well developed subgiant branch and a Red Clump are 
also seen. By fitting isochrones to this CMD as well as to other
CMDs available in the literature, and using the Red Clump location, the  
reddening, distance and age of the star cluster have been determined.
The cluster has a distance of $\sim$ 3.3 kpc, its radius is about 2.4 pc; 
the reddening E(B-V) is 0.08 mag and the age is $\sim$ 6.3 Gyr. By comparison 
with theoretical isochrones, a metallicity of [Fe/H] $\approx -0.2$ dex 
has been estimated.

Theoretical isochrones have been used to convert the observed cluster 
luminosity function into a mass function  in the mass range $\sim$ 0.6$-$1.1 $\Msun$. We find a much flatter mass function than what has been found for young clusters.
If the mass function is a power law $dN \sim m^{\alpha} dm$, then we get
 $\alpha = -0.5 \pm 0.3$.
This may be seen as a signature of the highly evolved dynamical state of the cluster.

\medskip
\parindent=0cm
{\bf Key words:} Open star clusters: individual: Berkeley 32 - star: evolution - HR diagram -
 Mass functions - Galactic disk

\end{abstract}
\section{Introduction}
Berkeley 32 ($C0655+065\sim$ OCL 522, l=207.$^{\circ}$95, b=4.$^{\circ}$4), also
known as Biurakan 8, is a small galactic (open) star cluster of angular
diameter $\sim 6{'}$. It is located in the Galactic anticentre direction and 
has been classified as Trumpler class II2m (Lyng\aa\ 1987). This object was 
discovered by Iskudarjan (1960) and catalogued by Setteducati \& Weaver (1960). 
On the sky survey maps, the cluster appears to be rich and likely of old age 
(cf. King 1964). The first photometric study of the cluster was carried 
out by Kaluzny \& Mazur (1991) in the $UBV$ and Washington systems by CCD 
imaging of an area $\sim 6.^{'}6 \times 6.^{'}6$. They presented the 
colour-magnitude diagram (CMD) and discussed its morphology, derived cluster 
reddening and distance as $E(B-V) = 0.16$ mag and 3.1$\pm$0.2 kpc respectively,
and estimated the metallicity as [Fe/H] = $-0.37\pm0.05$. They derived an age 
of $\sim$ 6 Gyr. Using morphological age parameters, Janes \& Phelps (1994)
estimated an age of 7.2 Gyr for Berkeley 32.

Scott et al. (1995) have determined radial velocities of 7 
cluster members and found a mean heliocentric radial velocity of +101$\pm$3 
km/sec, which among their cluster sample deviates most from a pure Galactic
rotation after Berkeley 17. 

Mass function studies of open star clusters indicated that the slopes of the 
mass functions of older (age $>$ 1 Gyr) clusters differ significantly from each 
other and also are not uniform over the entire mass range (cf. Sagar \& 
Griffiths 1998b). However, the number of old objects studied so far is small. 
As such, a study of the old cluster Berkeley 32 can contribute interesting 
knowledge on the mass function of old, probably highly evolved clusters. 

Our aim is to re-analyse Berkeley 32 with deeper photometry than Kaluzny \& 
Mazur (1991) had at their disposal and study its mass function which is 
lacking. The observations, data reductions and comparison with earlier 
photometry are given in the next two sections. The cluster radius,  other 
photometric results and mass function of the cluster are described in the 
subsequent sections of the paper.
\section{\bf Observations and Reductions }
The CCD observations have been obtained on 1999 March 22 with the 3.5m telescope
at Calar Alto Observatory, Spain, run by the Max-Planck Institute for astronomy,
Heidelberg. The focal reducer MOSCA attached at the RC Cassegrain focus 
provided an effective f/2.7 focal ratio 
({\bf http://www.caha.es/caha/instruments/mosca/manual.html}). The observations 
have been carried out in the Johnson $V$ and Cousins $I$ filters using the SITE 
18b CCD chip. Each pixel (24$\mu$ square) corresponded to $0.^{''}53 \times 
0.^{''}53$  on the sky. The non-vignetted area of the CCD was $ 1500 \times 
1500 $ pixel$^2$ providing a field of about 13.$^{'}$25 $\times$ 13.$^{'}25$. 
The read-out noise was 5.4 electrons per pixel and the ratio electrons-per-ADU 
was $\sim$ 2.7. Figure 1 shows the finding chart for the imaged cluster region 
and Table 1 lists the log of the observations. For calibration purpose, we 
observed the standard star field SA 98 (Landolt 1992). 

The present observations were carried out as a back-up programme. We therefore 
could not observe standard stars during the whole night. Instead, we observed 
the standard field SA 98 three times: before, between, and after the 
observations of Berkeley 32 (Table 1). The strategy was to observe Berkeley 32 
at a similar air-mass as the standard fields, so that one can calibrate 
photometric data of the cluster region with an accuracy of a few percent 
without a precise determination of the atmospheric extinction coefficients.
\begin{table}
{\bf Table 1.}~Log of CCD observations of the cluster Berkeley 32 and the
calibration region SA 98 (Landolt 1992). The data have been obtained on March 
22, 1999. 
\begin{center}
\begin{tabular}{ccc cc} \\ \hline 
Object    &   Time  &   Filter &  Exposure time & Air-mass \\
          &   (UT)  &          & (seconds)  \\ \hline
SA 98     & 19:23:45 &  $V$      & ~~8        &  1.27 \\ 
SA 98     & 19:29:00 &  $I$      & ~~8        &  1.27 \\
Be 32     & 19:40:04 &  $V$      & ~10        &  1.17 \\
Be 32     & 19:44:57 &  $V$      & ~60        &  1.18 \\
Be 32     & 19:49:05 &  $I$      & ~~8        &  1.18 \\
Be 32     & 19:51:47 &  $I$      & ~40        &  1.18 \\
Be 32     & 19:54:36 &  $I$      & ~~3        &  1.19 \\
SA 98     & 20:00:04 &  $V$      & ~10        &  1.30 \\
SA 98     & 20:02:37 &  $I$      &~~8         &  1.30 \\
Be 32     & 20:05:29 &  $V$      & ~10        &  1.20 \\
Be 32     & 20:07:44 &  $V$      & 100        &  1.20 \\
Be 32     & 20:11:37 &  $I$      & ~~8        &  1.21 \\
Be 32     & 20:14:04 &  $I$      & ~80        &  1.21 \\
SA 98     & 20:43:19 &  $V$      & ~10        &  1.39 \\
SA 98     & 20:45:37 &  $I$      & ~~8        &  1.40 \\ \hline
\end{tabular} 
\end{center}
\end{table}
The basic processing of the data frames was done in the standard manner using 
the MIDAS data reduction package. The uniformity of flat fields is better 
than one percent in both filters. 

Instrumental magnitudes have been measured using the DAOPHOT software (Stetson 
1987, 1992) under MIDAS. The image parameters and errors provided by DAOPHOT 
were used to reject poor measurements. About 10\% of the stars were rejected in 
this process. In those cases where brighter stars are saturated on deep exposure
frames, their magnitudes have been taken only from the short exposure frames. 
Most of the stars brighter than V $\sim$ 12 mag could not be measured because 
they were saturated even on the shortest exposure frames. 
\begin{table}
{\bf Table 2.}~Colour coefficients and zero-points are for 1 second exposure 
time of the standard stars. 
\begin{center}
\begin{tabular} {ccc cc } \\ \hline
 Air-mass &  a0$\pm \sigma$  &  a1$\pm \sigma$& b0$\pm \sigma$  & b1$\pm \sigma$ \\ \hline
1.27  & $-0.513\pm0.01$ & $0.07\pm0.01$ & $0.667\pm0.02$ & $0.03\pm0.003$ \\
1.30  & $-0.523\pm0.02$ & $0.07\pm0.01$ & $0.664\pm0.02$ & $0.03\pm0.003$ \\
1.39 & $-0.573\pm0.02$ & $0.07\pm0.01$ & $0.643\pm0.02$ & $0.03\pm0.003$ \\
\hline
\end{tabular} 
\end{center}
\end{table}
The CCD instrumental magnitudes have been calibrated using the observations of 
the SA 98 field and the following relations 
\[ V-v = a0 + a1*(V-I); \hspace{2cm}  (V-I) - (v-i) = b0 +b1*(V-I)\]
where capital letters denote standard magnitudes and colours, and lower case 
letters denote instrumental values. The values refer to exposure time of 
1 second. These equations along with the site mean atmospheric
 extinction values of 0.15$\pm$0.04 and 0.09$\pm$0.02 mag per unit air-mass in 
$V$ and $(V-I)$ respectively were used in determining the colour equations for 
the system as well as the zero-points. The effects of uncertainties in 
atmospheric extinction values are maximum on zero-points but least on the 
colour coefficients. We therefore averaged the colour coefficients from the 
individual standard observations. With these values fixed we calculated the 
zero-points. Table 2 lists the colour coefficients and zero-points derived in 
this way. The zero-points are uncertain by $\sim$ 0.02 mag in $V$ and $(V-I)$. 
The internal errors as a function of magnitude for each filter are given in 
Table 3. The errors become large ($>$ 0.15 mag) for stars fainter than $V = 22$ 
and $I = 21$ mag. 
\begin{table}
{\bf Table 3.}~Internal photometric errors as a function of brightness.
$\sigma$  is the standard deviation ($\sigma$) per observations in magnitude.
\begin{center}
\begin{tabular}{ccc}\\ \hline  
 Mag range & $\sigma_V$ & $\sigma_I$  \\ \hline
 12.0 $-$ 14.0 & 0.005 & 0.010  \\
 14.0 $-$ 16.0  &0.005 & 0.010  \\
 16.0 $-$  17.0   &0.005 & 0.010  \\
 17.0 $-$  18.0   &0.006 & 0.013  \\
 18.0 $-$  19.0   &0.009 & 0.024  \\
 19.0 $-$  20.0   &0.017 & 0.051  \\
 20.0 $-$  21.0   &0.041 & 0.117  \\
 21.0 $-$  22.0   &0.106 &   \\
 22.0 $-$  23.0   &0.221 &   \\ \hline
\end{tabular} 
\end{center}
\end{table}
The $X$ and $Y$ pixel coordinates as well as the $V$ and $(V-I)$ magnitudes 
and DAOPHOT errors of the stars observed in Berkeley 32 are listed in Table 4.  
Stars observed by Kaluzny \& Mazur (1991) have been identified in the 
last column. Table 4 is available only in electronic form at the open cluster
database Web site at {\bf http://obswww.unige.ch/webda/.} It can also be 
obtained from the authors.

\section  {\bf Comparison with previous photometry}

   In this section, we compare the present CCD photometry with the only 
previous CCD photometric observations of the cluster by 
Kaluzny \& Mazur (1991) in the only common passband $V$. The transformation
equations relating their ($X_{km}, Y_{km}$) coordinate system
to ours ($X_{pres}, Y_{pres}$) were found to be 
\[  X_{pres} = 1178.252 - 0.024 X_{km} - 1.512 Y_{km}; \hspace{1cm}
  Y_{pres} = 259.383 + 1.511 X_{km} - 0.024 Y_{km} \]
There are 835 stars measured by Kaluzny \& Mazur (1991) whose positions coincide
within 1 pixel with the stars positions measured by us. The differences 
($\Delta V$) between the present data and data obtained by them are plotted in 
Fig. 2,  while the statistical results are given in Table 5. These show that 
except for a few outliers, which appear to be mostly stars that were treated as 
single in one study and as double (due to blending) in the other, the 
distribution of the photometric differences seems fairly random with almost no 
zero-point offset. As expected, the scatter increases with decreasing 
brightness and becomes more than $\sim$ 0.1 mag at fainter levels. Considering 
the uncertainties present in our and Kaluzny \& Mazur's (1991) measurements, 
we conclude that they are in very good agreement. 
\begin{table}
{\bf Table 5.}~Statistical results of the photometric comparison with data 
from Kaluzny \& Mazur (1991). The difference ($\Delta$) is in the sense 
present minus comparison data. The mean and standard deviation ($\sigma$) 
are based on N stars. A few points discrepant by more than 3.5 $\sigma$ 
have been excluded from the analysis.
\begin{center}
\begin{tabular}{ccc ccc} \\ \hline 
$ V$ range &\m{2}{c}{$\Delta V$}&$(V-I)$ range&\m{2}{c}{$\Delta V$} \\
    (mag) & Mean$\pm \sigma$ & N  &(mag) & Mean$\pm \sigma$ & N  \\ \hline
 12$-$14 & $~0.006\pm$0.03 & 17&$-0.1$$-$0.65 & $-0.021\pm$0.16 &  41  \\ 
 14$-$16  &$-0.012\pm$0.04 & 67&0.65$-$0.8~  &   $-0.013\pm$0.12  & 197 \\
 16$-$17   &  $-0.004\pm$0.04 &129&0.8~$-$1.0~  &  $-0.020\pm$0.12   & 242 \\
 17$-$18   &  $-0.015\pm$0.08 & 144&1.0~$-$1.5~  &  $-0.026\pm$0.11   & 286 \\
 18$-$19   &  $-0.029\pm$0.09 & 159&1.5~$-$3.2~ & $ -0.011\pm$0.11  &  69  \\ 
 19$-$20   &  $-0.015\pm$0.14 & 157 \\
 20$-$21   &  $-0.023\pm$0.16 &  162 \\ \hline
\end{tabular} 
\end{center}
\end{table}
\section {Radius of the cluster}

We used radial stellar density profile for the determination of cluster radius. 
Such determinations can provide ambiguous results  as it depends on the 
limiting magnitude of the star counts. The fainter the stars are, the larger 
becomes the cluster radius, if mass segregation due to two-body relaxation is 
present. Given these caveats, it is not our aim to derive a dynamically relevant
radius, but to determine the region where the cluster population dominates over
field stars so that it can be used for investigations of the cluster properties.  

We derived the position of the cluster centre by iteratively 
calculating the average X and Y positions of stars within 150 pixels from an 
eye estimated centre, until it converged to a constant value. The (X,Y) pixel 
coordinates of the cluster centre are (700, 665) with an accuracy of few pixels.
The radial stellar density profile determined up to $\sim 6^{'}$ from the 
cluster centre using stars brighter than V=18 mag is plotted in Fig. 3. The 
radius at which the star density flattens is considered as cluster radius which 
is $\sim 2.^{'}7$. This agrees fairly well with the value of 3$^{'}$ given by 
Lyng\aa\ (1987). We fit the following form of a 
King (1962) profile to the observed stellar density distribution 

\[f(r)\propto C\cdot (\frac{1}{\sqrt(1+(r/r_c)^2)}-\frac{1}{\sqrt(1+(r_t/r_c)^2)})^2,\]

\noindent where $C$ is the central stellar density, $r_c$ and $r_t$ are the 
core and tidal radius respectively. A least square fitting of the profile to 
the observed points yielded  $C = 33.9 \pm 8$ star/arcmin$^2, r_c = 1.0\pm 0.38,
r_t = 23 \pm 50$ arcmin.

\section {Colour-magnitude diagrams (CMDs)} 
\subsection{The $V, (V-I) $ CMD}
We plot the $V, (V-I)$ CMD for all ($\sim$ 3200) measured stars in the 
region of Berkeley 32 in Fig.4(A). The CMD reaches down to $V = 22$ mag. The 
cluster main-sequence (MS) contaminated by field stars is clearly visible. 
Although it is clear that the stellar population of this region is of composite 
nature, the cluster population appears to be dominating. The only way to 
sharpen morphological features of the cluster sequence in the CMD is to select 
stars with small radial distances by compromising between a decreasing 
number of cluster stars and an increasing field population. The Fig. 4 (B) 
shows our best result. Here we have selected only stars with a radial distance 
up to $\sim 2.^{'}7$. The features of a very old open star cluster namely the 
distinct turn-off region and the subgiant branch are now very clearly visible. 
The giant branch (GB) is very sparsely populated and not well defined. 
Moreover, a group of stars can be seen which are brighter and bluer than the MS 
turn-off point suggesting that some of them are blue stragglers (BSs). Such 
objects have been found in most intermediate and old age open star clusters (see
Kaluzny 1994; Phelps et al. 1994; Sagar \& Griffiths 1999a). Many of them have 
been identified as close binary systems. The (X,Y) pixel coordinates, radius, 
magnitudes and colour of the stars located in the GB, red GB and BS regions of 
the CMDs are given in Table 6. The cluster membership of these stars is also
indicated in the table. A star is considered as probable cluster member if it 
lies within $\pm$0.05 mag in colour and $\pm$0.1 mag in brightness with respect 
to the isochrone of the cluster age at least in two of the $V, (U-V); V, (B-V)$ 
and $V, (V-I)$ diagrams. In addition, brightening due to unresolved/optical 
binary stars has also been considered.

Fig. 4 (C) shows the $V, (V-I)$ diagram of stars with radial distances more
than $\sim 4.^{'}4$ from the cluster centre. Overplotted are the fiducial
points of the cluster sequence. There are a few red giants, which perhaps 
still belong to the cluster. A considerable part of the main sequence population
has a turn-off similar to the cluster, but the bulk of the main sequence stars 
are clearly shifted towards the red, indicating higher reddening, and thus a 
background population. However, the interesting question whether there are 
evaporated stars surrounding the cluster can not be answered on the basis of 
the present data. For this, kinematic informations like proper motions and 
radial velocities of these stars are required.
\subsection {The cluster age from the "Red Clump"}
It is well known that for intermediate and old open star clusters, the location
of the Red Clump (RC) (the more massive analog of the horizontal branch in 
globular clusters) relative to the MS turn-off point is correlated with age 
(cf. Kaluzny 1994; Phelps et al. 1994; Carraro \& Chiosi 1994; Pandey et al. 
1997 and references therein). The two morphological parameters generally used 
for estimating cluster ages are the differences in magnitudes ($\triangle V$) 
and colours ($\triangle (B-V)$ or $\triangle (V-I)$) between the RG branch at 
the level of the clump and the MS turn-off point, with the advantage that no 
prior knowledge of cluster distance, reddening and accurate metallicity is required.

Following Kaluzny (1994), we find $\triangle V = 2.7\pm0.05, \triangle (V-I) = 
0.45\pm0.03$ in the case of Berkeley 32. Using the relation given by Carraro \& 
Chiosi (1994), we derive log (age) = $9.8\pm0.1$ for the cluster.
A slightly modified version of $\triangle V$ has been introduced by Janes \& 
Phelps (1994) who used the luminosity difference between the RC and the 
inflection point between the turn-off region and the subgiant branch. We 
confirm their value of 2.4 for Berkeley 32. From their Fig.1 one reads off the 
logarithm of age as 9.8-9.9, yielding an age of 6.3 - 8 Gyr. According to their 
Table 1, there are only a few clusters older than Berkeley 32, like NGC 6791, 
Berkeley 54, AM 2 and Cr 261. It is thus clear that Berkeley 32 
belongs to the group of very old open clusters in our Galaxy.
\subsection{Determination of the cluster parameters using theoretical isochrones}  
We have determined the colour excess, the distance modulus, and also the age 
of the cluster by fitting theoretical stellar evolutionary isochrones 
from the set of Bertelli et al. (1994) to our $V, (V-I)$ diagram.  These
isochrones are derived from stellar models computed with updated
radiative opacities and include the effects of mass loss and
convective core overshooting. The models trace the evolution from the 
zero-age main-sequence (ZAMS) to the central carbon ignition for massive stars
and to the beginning of the thermally pulsing regime of the asymptotic giant
branch phase for low and intermediate mass stars.

As most of the factors responsible for the colour spread in the MS will
redden the stars (differential reddening, binaries, rotation, star spots),
we have used the blue envelope of the MS in the CM diagram for the estimation 
of the cluster parameters. We fit the isochrones by eye taking into account the 
observational error. It turns out that the isochrone with log(age) = 9.8, $X = 
0.7, Y = 0.28$ and $Z = 0.008$ fits best to the cluster locus, including the RC,
and thus is in good agreement with the age estimated from the morphological 
parameters of the cluster CMD. In order to also demonstrate upper limits of the 
effects of binaries in the CMD, the log (age) = 9.8 isochrone for the single 
stars has been brightened by 0.75 mag leaving the colour unchanged. A maximum  
reddening of $E(V-I) = 0.11$ (or $E(B-V) = 0.08$) mag can be applied to place 
the isochrone correctly on the cluster sequence observed in Fig. 4(B). Some 
stars above the turn-off point lie on the isochrones of binaries indicating 
the possibility of being indeed binary members of the cluster.
The lower giant branch in the $V, (V-I)$ diagram appears marginally too blue, 
indicating that the cluster may have a slightly higher metallicity. However,
a $Z=0.02$ isochrone is definitely too metal-rich. Moreover, a solar 
metallicity would decrease the cluster reddening even further, while Kaluzny \& 
Mazur (1991) quote a reddening of $E(B-V) = 0.16$ mag. However, such a high 
reddening is supported only by their $V, (U-V)$ diagram (see section 5.4).

It can also be seen that the  theoretical location of the RC fits rather well 
with the observed  one for Berkeley 32  unlike in some other old open star
clusters (see Sagar \& Griffiths 1999a). For example, it is too faint for 
NGC 6603 and too bright for NGC 7044.

The value of the apparent distance modulus derived from Fig. 4(B) is 12.8 mag. 
Here we adopt the reddening law of Rieke \& Lebofsky (1985), who give 
$A_V/E(V-I) = 1.94$. With the extinction $A_V = 0.21$ mag, we get for Berkeley 
32, a true distance modulus of 12.6 with an uncertainty of $\sim$ 0.15 mag 
which includes errors in the photometric calibration, isochrone fitting and 
the reddening determination.  
\subsection{Isochrone fitting to the UBV data of Kaluzny \& Mazur (1991)}
Fig. 5 shows the $V, (U-V)$ and $V, (B-V)$ diagrams generated from the
 Kaluzny \& Mazur (1991) photometric data. Overplotted are the isochrones of 
Bertelli et al. (1994), having the same ages, metallicity and helium 
abundance as we used for our $V, (V-I)$ diagram in Fig 4(B). In order to fit 
the isochrone to the cluster sequence, we had to employ a reddening of 
$E(B-V) = 0.08$ mag and $E(U-V) = 0.22$ mag in the $V, (B-V)$ and $V, (U-V)$ 
diagrams respectively. While for the reddening law of Rieke \& Lebofsky (1985),
the $E(B-V)$ value agrees well with our E(V-I) value ($E(V-I)/E(B-V)$ = 1.6), 
the $E(U-V)$ is too large ($E(U-V)/E(B-V)$=1.64). On the other hand, it is too 
small for $E(B-V) = 0.16$ mag, given by Kaluzny \& Mazur (1991) and is thus not 
compatible with the reddening values derived from the $V, (V-I)$ and $V, (B-V)$ 
diagrams. This may suggest that the $U$-photometry is perhaps in error and we 
adopt $E(B-V) = 0.08$ mag as the value for the cluster reddening. 
\subsection{The cluster distance from the Red Clump}
For a star cluster as old as Berkeley 32, an attractive method to determine its
distance is using the location of the RC of intermediate-age helium core
burning stars as a standard candle (e.g. Paczynski \& Stanek 1998).
The absolute I-magnitude of RC stars in the solar neighborhood has been
calibrated by Hipparcos parallaxes, resulting in $M_I^0 = -0.23 \pm0.03$.
Cole (1998) discusses the age and metallicity dependence of the RC-brightness
and notes that for poulations older than 4-5 Gyr, the $M_I^0$ is independent
of stellar mass, but still shows a metallicity dependence of the order $\delta
M_I^0 = (0.21\pm0.07) [Fe/H]$, where $\delta M_I$ is the brightness difference 
between the RC in the solar neighborhood and the population under consideration.
As the value is small for Berkeley 32, we neglect this correction here.     

The RC in Berkeley 32 has $V=13.67\pm0.03$ and $(V-I) = 1.16\pm0.03$, where the 
error is the uncertainty in the definition of the RC in the CMD. This yields
$(m-M{_I}) = 12.74 \pm0.08$ as the apparent distance modulus, if we include the 
photometric calibration uncertainty in the error. The extinction in $I$ is 
determined by $A_I/E(V-I) = 0.93$ which is 0.10 for a value of $E(V-I) = 0.11$. 
If we assign an additional error of 0.05 to the extinction correction, we have 
$(m-M)_0 = 12.64\pm0.1$ as the value for true distance modulus. This agrees well
with the value obtained using isochrone fitting. In the following, we therefore
adopt $12.6\pm0.1$ as the  value for the distance modulus of the cluster.
The present distance determination of 3.3$\pm$0.2 kpc agrees very well with 
the value of 3.1 kpc given by Kaluzny \& Mazur (1991). 
The cluster parameters derived by us are listed in Table 7. 
\begin{table}
{\bf Table 7.} The age, metallicity $([Fe/H])$, reddening ($E(B-V)$ and 
$E(V-I)$), true distance modulus($(m-M)_0$), distance, galacto-centric distance 
($R_{GC}$) (adopting galacto-centric distance of the Sun as 8 kpc) 
and z-distance for Berkeley 32.
\begin{center}
\begin{tabular}{ll} \hline
Age &  6.3 Gyr  \\
$[Fe/H]$ & -0.2 dex \\
$E(V-I);E(B-V)$ & 0.11;0.08\\
$(m-M)_0$ & $12.60\pm0.1$ \\
Distance  & 3.3 kpc \\ 
$R_{GC}$ & 11.0 kpc \\
z & 250 pc \\ \hline
\end{tabular}
\end{center}
\end{table}
\subsection {Location of Berkeley 32 in the Galaxy}
The cluster Berkeley 32 occupies an important position for understanding the
variation of metallicity in the Galactic disk, as the issue of the existence 
of a metallicity gradient is not yet settled. According to 
Friel (1995), the metallicities of open clusters indicate a gradient of $-0.09$
dex/kpc. On the other hand, Twarog et al. (1997) argue that the open cluster 
system can be divided in 2 radial groups, with a very flat or even vanishing 
gradient in each group. Their mean metallicities differ by 0.3 dex and there 
is a discontinuity at a radial distance of 10  kpc. As the galactocentric 
distance of Berkeley 32 puts it just near this discontinuity, Berkeley 32 
could help to decide between these two  metallicity patterns in the 
Galactic disk. However, a more accurate determination of the metallicity than 
we are able to do, is required. Also, more clusters/objects either in the 
vicinity of  Berkeley 32 or at similar galacto-centric distances need to be 
observed before the the metallicity pattern can be unambiguously determined. 
\section{Mass function}
The study of the mass function (MF) of Berkeley 32 is based on a pair of deep 
$V$ and $I$ CCD frames only. This is done for evaluating the data completeness 
accurately. Guided by the radial stellar density profile in Fig. 3, we selected 
stars located within a circle of 165 arcsec radius (surface area 23.76 square 
arcmin) around the cluster centre for the MF study. With the aim of detecting  
possible radial MF variations, the data completeness has first been evaluated 
in an inner and outer region separately, but it turned out to be the 
same within the errors. Besides, the small number statistics prevented us from
a detailed study. We therefore determined the MF for the entire region without 
any subdivision.   

To suppress the field star contamination as far as possible,
we selected cluster main sequence stars by
 using the following boundaries in the $V, (V-I)$ diagram: 
$$ V > 8.91+9.45 \cdot (V-I) + 1.8 \cdot (V-I)^2\\$$
and\\
$$ V < 11.52 + 12.64 \cdot (V-I) + 3.86 \cdot (V-I)^2\\$$ 
The field star contamination has been determined using the remaining chip area 
outside a radius of 265 arcsec (surface area 123.1 square arcmin) from the 
cluster centre. Table 8 lists the field  and cluster counts derived in this way 
along with their completeness factors. The completeness factors have been 
determined by using artificial stars along the clusters main sequence and 
recovering them in the CMD, not in one filter, to be as realistic as possible.
\begin{table}
{\bf Table 8.} The V-magnitude of the bin center, the raw counts for cluster 
($N_C$) and field regions ($N_F$), and the corresponding completeness factors 
$f_C$ and $f_F$ are listed. Absolute $M_V$ and stellar mass (m) of the bin 
center, the mass interval ($\Delta m$) corresponding to the magnitude bin, the 
normalised counts ($N$) and their errrors ($\delta N$) are also listed.
\begin{center}
\begin{tabular}{lllll lllll} \hline
V&$N_C$&$N_F$&$f_C$&$f_F$&$M_V$&m&$\Delta m$&$N$&$\delta N$\\ \hline
16.25& 41&  53& 1.00& 1.00   & 3.40 &1.113& 0.0693& 18.70&  4.2 \\
16.75& 44&  67& 1.00& 1.00   & 3.90 &1.046& 0.0658& 19.88&  4.6 \\
17.25& 47&  86& 1.00& 1.00   & 4.40 &0.982& 0.0623& 20.54&  5.0\\
17.75& 43& 107& 1.00& 1.00   & 4.90 &0.921& 0.0588& 16.00&  5.1 \\
18.25& 50& 135& 0.95& 1.00   & 5.40 &0.864& 0.0553& 20.22&  6.3\\
18.75& 37& 121& 0.85& 1.00   & 5.90 &0.811& 0.0518& 16.39&  6.3\\
19.25& 35& 127& 0.80& 0.93   & 6.40 &0.761& 0.0483& 15.15&  7.4\\
19.75& 31& 158& 0.81& 0.89   & 6.90 &0.714& 0.0449&  3.76&  7.4\\
20.25& 37& 154& 0.80& 0.90   & 7.40 &0.671& 0.0413& 13.46&  8.8\\
20.75& 31& 142& 0.73& 0.86   & 7.90 &0.631& 0.0379& 11.73&  8.9\\
21.25& 24& 116& 0.60& 0.70   & 8.40 &0.595& 0.0344&  9.82&  11.1\\ \hline
\end{tabular}
\end{center}
\end{table}
Table 8 also lists the numbers which are relevant for the MF determination. The 
transformation from apparent to absolute visual magnitude ($M_V$) has been 
done using the cluster parameters given in Table 7. The isochrone log(age) = 9.8
and $z=0.008$ provides the following parametrization of mass $(m)$ and $M_V$:
$$ m = 1.665 -0.186 \cdot M_V + 0.00698 \cdot M_V^2.$$
The values of the normalised counts ($N$) are in stars/arcmin$^2$. They are 
corrected for completeness and field star contamination, and divided by 
the mass interval of the magnitude bin. The errors of the normalised counts 
result from error propagation. This may be incorrect from a puristic viewpoint, 
as they are no longer small compared to the counts. However,  we do 
not use them any further beyond a qualitative demonstration that they are large.

A common description of the stellar mass spectrum is a power law $dN \propto
 m^\alpha dm$, where $dN$ is the number of stars in the mass interval $m+dm$,
 and $\alpha$ is the slope of the MF. However, a 
uniform exponent is at best realised within limited mass intervals. The 
universality of the slope of the initial mass function (IMF) is still a 
matter of discussion (for a recent review see Scalo 1998), but studies of 
a large number of young clusters in the Milky Way and the Large Magellanic
Clouds do not speak evidently against an universal IMF at least above 1 
$M_{\odot}$ (e.g. Sagar et al. 1986; Sagar \& Richtler 1991; Janes \& Phelps 
1994; Fischer et al. 1998; Sagar 2000), with $\alpha$ around $-2.3$. 
Below 1 $M_{\odot}$, the data for young open clusters are sparse and any 
secure statement on a possible universal IMF is not yet possible. 
 
However, in the case of a very old cluster like Berkely 32, we might anyway not 
 expect the 
IMF to be still realized. As a cluster evolves dynamically, low mass stars 
evaporate out of the cluster potential faster than high mass stars. In a cluster
 much older than its 
relaxation time, the dynamical effect therefore can change an originally rising 
IMF into a flat or even declining MF. 

Fig. 6 shows the MF for Berkeley 32. The logarithm of mass is 
plotted against the logarithm of the normalised counts. Note that the binning
in mass is also logarithmic. Although the errors are too large for any deeper 
analysis, it is apparent that the MF is much flatter than of most young 
clusters. A fit to a power-law indeed returns the value $\alpha = -0.5 \pm 0.3$
while we expect in this mass domain an exponent around $-2$ for young clusters 
(Richtler 1994). Here it must be remarked that we assumed single stars  for
the applied mass-luminosity relation of the isochrone. Sagar \& Richtler (1991)
discussed how the presence of binaries flattens a "true" IMF. But even if we 
find a large binary fraction in Berkeley 32, as the CMD suggests, their effect 
would by far not be sufficient to steepen the observed MF to the level as it 
is observed for young clusters. 

This behaviour, in agreement with theoretical expectations, has been found for 
other old open clusters as well. For example, Francic (1989), among his
sample of 8 open clusters, found the old objects NGC 752 (2.5 Gyr) and M67 
(5 Gyr) to show even declining mass functions with $\alpha > 0$. Further well 
studied open clusters like NGC 6791 (Kaluzny \& Rucinski 1995), NGC 188 (von 
Hippel \& Sarajedini 1998), NGC 2243 (Bergbusch et al. 1991) also have flatter 
MFs. But one also can find old clusters with MF not distinguishable from a 
Salpeter mass function, e.g., mass spectrum of Berkeley 99 (age 3.2 Gyr) has 
$\alpha \sim -2.4$ (Sagar \& Griffiths 1998b). This demonstrates that open 
clusters do have distinctly different dynamical histories, which may depend 
on their structure, total mass, location, orbit characteristics etc. 
\section{Summary}
New $V$ and $I$ CCD photometry down to $V = 22$ mag is presented for about 3,200
stars in the region of the open cluster Berkeley 32. The present photometry
serves as a data base for determining the cluster properties and to study the 
stellar mass function for the first time. The cluster's radial density profile 
is well represented by a King (1962) profile. By fitting of theoretical 
isochrones and using the location of the Red Clump, we confirm earlier results 
that it is indeed a very  old open cluster (6.3 Gyr). Its  metallicity is 
between $Z = 0.008$ and $Z = 0.02$, distance is 3.3 kpc and galacto-centric 
distance is 10.8 kpc. Clusters/objects with these characteristics can play 
a very valuable role to distinguish  between the two models of the metallicity 
variation in the Galactic disk, advocated by Friel (1995) and Twarog et al. 
(1997) respectively. However, the case of Berkeley 32 is ambiguous. The 
parameters of Berkeley 32 are compatible with both a smooth Galactic 
metallicity gradient as well as with its membership of the cluster population 
of the inner domain of Twarog et al. (1997).

We also investigated the mass spectrum of Berkeley 32 in the mass range 
0.6-1.1 $M_{\odot}$. A power-law fit returns $\alpha = -0.5\pm0.3$ for the 
slope of the MF, which is much flatter than the slopes found in young open 
clusters. Berkeley 32 shares this behaviour with other old open clusters 
which indicates an evaporation of low-mass cluster stars.

\vspace{0.5cm}

{\bf ACKNOWLEDGEMENTS}

The suggestions/comments given by the referee Randy L. Phelps improved the
 presentation and readablity of the paper. 
 RS gratefully acknowledges the support from the Alexander von Humboldt 
 Foundation.  TR wants to thank the U. P. State Observatory, Nainital, the  
 Deutsche Forschungsgemeinschaft, and the Indian National Science Academy, for 
 financial support and warm hospitality. Thanks to Klaas de Boer for going 
 through the manuscript critically. The {\bf Open Cluster Data Base} 
 maintained by J.-C. Mermilliod has been used in the present work.

\bigskip

{\bf REFERENCES}

\begin{description}
 \item [ ] Bergbusch P. A., VandenBerg D.A., Infante L. 1991, AJ 101, 2102
\item [ ] Bertelli G., Bressan A., Chiosi C., Fagotto F., Nasi E., 1994,
   A\&AS 106, 275
\item [ ] Carraro G., Chiosi C., 1994, A\&A 287,  761
\item [ ] Cole A.A., 1998, ApJ 500, L137
\item [ ] Fischer P., Pryor C., Murray S., Mateo M., Murray S., Richtler, T., 
1998, AJ 115, 592
\item [ ] Francic S.P., 1989, AJ 98, 888 
\item [ ] Friel E.D., 1995, ARA\&A 33, 381
\item [ ] Iskudarjan S.G., 1960, Comm. Biur. 28, 46
\item [ ] Janes K.A., Phelps R.L., 1994, AJ 108, 1773 
\item [ ] Kaluzny J., Rucinski S.M., 1995, A\&AS 114, 1
\item [ ] Kaluzny J., Mazur B., 1991, Acta Astr. 41, 167
\item [ ] Kaluzny J., 1994, A\&AS 108, 151
\item [ ] King I.R., 1962, AJ  67, 471
\item [ ] King I.R., 1964, Royal Obser. Bull. 82, 106
\item [ ] Landolt A., 1992 AJ 104, 340 
\item [ ] Lyng\aa\ G., 1987. {\it Catalogue of Open Cluster Data}, 5th edition
           , 1/1 S7041, Centre de Donnees Stellaires, Strassbourg.
\item [ ] Paczynski B., Stanek K.Z., 1998, ApJ 494, L219
\item [ ] Pandey A.K., Durgapal A.K., Bhatt B.C., Mohan V., Mahra H.S., 1997,
           A\&AS, 122, 111
\item [ ] Phelps R.L., Janes K.A., Montgomery K.A., 1994, AJ  107, 1079
\item [ ] Richtler T., 1994, A\&A 257, 517 
\item [ ] Rieke G.H.,  Lebofsky M.J., 1985,  ApJ 288, 618
\item [ ] Sagar  R.,  2000, BASI 28, 55
\item [ ] Sagar  R.,  Griffiths  W.K., 1998a, MNRAS 299, 1 
\item [ ] Sagar  R.,  Griffiths  W.K., 1998b, MNRAS 299, 777 
\item [ ] Sagar R., Richtler T., 1991, A\&A 250, 324
\item [ ] Sagar R., Piskunov A.E., Myakutin V.I., Joshi U.C., 1986, 
           MNRAS 220, 383
\item [ ] Scalo J., 1998, in The Stellar Initial Mass Function (38th Herstmonceux
 Conf.), eds. G. Gilmore 
            and D. Howell, ASP Conf.Ser. 142 (1998), p.201
\item [ ] Scott J.E., Friel E.D., Janes K.A., 1995, AJ 109, 1706
\item [ ] Setteducati A.E., Weaver M.F., 1960 in Newly found stellar clusters,
    Radio Astronomy Laboratory, Berkeley
\item [ ] Stetson P.B., 1987, PASP  99, 191
\item [ ] Stetson P.B., 1992, in Astronomical Data Analysis Software and 
 Systems I. Worrall D.M., Biemesderfer C. and Barnes J. (eds), 
 ASP Conf. Ser. 25, p. 297
\item [ ] Twarog B., Ashman, K.M., Anthony-Twarog B.J., 1997, AJ 114, 2556
\item [ ] von Hippel T., Sarajedini A., 1998, AJ 116, 1789 
\end{description}

\newpage
{\bf Figure captions}

\begin{description}

\item [Fig. 1] Identification chart for the Berkeley 32 region. The (X,Y) 
coordinates are in CCD pixel units and one CCD pixel corresponds to 0.$^{''}$53
on the sky.  North is up and East is to the left. Filled circles of different
sizes represent the brightness of the stars. The smallest size denotes stars 
of $V = 17$ mag.

\item [Fig. 2] Comparison of the present $V$ magnitude with those of Kaluzny 
\& Mazur (1991). The differences ($\triangle$) are in the sense of this study 
minus Kaluzny \& Mazur. They are plotted against the present CCD photometry. 

\item [Fig. 3] Plot of the radial density profile ($\bullet$) for stars 
brighter than $ V = 18$ mag in Berkeley 32 region. The length of the bar
represents errors resulting from sampling statistics. Overplotted (solid curve)
is a King (1962) profile with parameters given in the text. The arrow denotes 
the radius where the surface density of cluster stars becomes becomes 
comparable with the field star density.

\item [Fig. 4] The $V, (V-I)$ diagrams (A) for all stars observed by us, (B) 
for the cluster population (stars with radius $\le 2.^{'}7$) and (C) for the 
field population (stars with radius $\ge 4.{'}4$) in the Berkeley 32 region are 
plotted. The composite nature of the stellar population is apparent in (A).
In the cluster population, the Bertelli at al. (1994) isochrones 
for $z=0.008$ and log (age) = 9.7 (short-dashed), 9.8 (continuous) and 9.9 
(dot-short dashed) are shown. The isochrone of log (age) = 9.8 found to be 
best fitting to the observed cluster sequence with a reddening of $E(V-I) = 
0.11$ (or $E(B-V) = 0.08$) and an apparent distance modulus of 12.8 mag. The 
dotted curve shows the extent that binaries of equal mass can brighten the 
isochrone of log (age) = 9.8. 
In the field population, the overplotted curve is displaying the cluster locus. 
There are red giants resembling cluster giant branch stars. A considerable 
part of the main sequence population has a turn-off similar to the cluster, 
but the bulk of the main sequence stars are clearly shifted towards the red, 
indicating higher reddening, and thus a background stellar population.

\item [Fig. 5] The $V, (U-V)$ and $V, (B-V)$ diagrams generated for the 
cluster population of Berkeley 32 from the photometric data of Kaluzny \& 
Mazur (1991). The Bertelli et al. (1994) isochrones of the same metallicity and 
ages, shifted by the same value of the apparent distance modulus as in Fig. 
4(B) are shown. The best eye fits to the cluster sequence 
for the reddening values shown in the plot. The $E(U-V)$ value is not 
compatible with $E(B-V)$, indicating a problem with the $U$ photometry.
\item [Fig. 6]
The mass function of Berkeley 32 between 0.6 and 1.1 $M_\odot$ as derived from
the present data. Although the error bars are large, the mass function is 
clearly flatter than what has been found for young star clusters.
\end{description}
\newpage

\noindent {\bf Table 6.}~Spatial and $UBVI$ photometric values of the 
candidate giant
branch and blue straggler stars are listed along with the identification of
Kaluzny \& Mazur (1991) prefixed with KM. The $(U-B)$ and $(B-V)$ values are
taken from Kaluzny \& Mazur (1991). The probable photometric members 
have been identified as PM in the last column.

\scriptsize
\begin{tabular}{rrr rcc ccrc} \\ \\  \hline
\m{10}{c}{(A) Stars redder than MS turn-off (candidate for giant branch)} \\ \hline
Star &X & Y & Radius &$V$ &$(U-B)$ & $(B-V)$ & $(V-I)$ & Other &Membership \\
  &(pixel) &(pixel) &(pixel)&(mag)&(mag)&(mag)&(mag)&identification \\ \hline
   226 & 813.65&  509.20&  192.85& 16.00& 0.18 &0.64 & 0.74& KM86 &  PM \\
   238 & 525.10 & 526.47& 223.12 & 15.57& 0.83 &1.05 & 1.14& KM59 &   \\
   242&  667.64 & 532.82&  136.08& 13.86&0.85 & 1.10 & 1.19& KM21 &  PM \\
   245&  720.59&  535.05&  131.57& 16.22&0.07 &0.59  & 0.71& KM120 &  PM \\
   254&  679.15&  546.71&  120.11& 15.87&0.62 &0.97 & 1.10& KM74 & PM \\
   264&  426.71&  564.67&  291.12& 16.03&0.15 &0.67 & 0.79 & KM108 & PM \\
   269&  975.20&  571.96&  290.50& 14.28&0.88 &1.14& 1.25& KM24 & PM   \\
   274&  947.00&  584.20&  259.88& 13.27&0.66 &1.01 & 1.13& KM10 &   PM \\
   293&  760.87&  617.50&   77.21& 16.08&0.11 &0.62 & 0.74& KM103 &   PM \\
   302&  603.93&  629.92&  102.27& 16.02&0.09 &0.57 & 0.69& KM91 &   PM \\
   310&  498.03&  639.10&  203.62& 15.28&0.80 &1.01 & 1.06& KM46 &   PM \\
   318&  880.21&  657.52&  180.37& 13.76&0.78 &1.08 & 1.18& KM19 &   PM \\
   331&  584.61&  683.34&  116.84& 16.14 &0.30 &0.77 & 0.88& KM113 &   PM \\
   332&  682.86&  684.53&   25.98& 16.00&1.10 &1.15 & 1.34&KM93  &   \\
   347&  748.92&  706.51&   64.16& 12.90&0.68 &1.02 & 1.06& KM8 &   \\
   352&  514.50&  720.01&  193.48& 14.98&0.12 & 0.68  & 0.78& KM38 &   \\
   357&  812.84&  727.41&  128.95& 15.88&0.63 &0.96  & 1.04& KM75 &   PM \\
   360&  534.13&  733.92&  179.62& 14.76&0.80 &1.16  & 1.25& KM33 &   \\
   414&  759.17&  809.22&  155.89& 16.01&0.19 &0.69 & 0.82& KM87 &   PM \\
   416&  904.29&  812.04&  251.70& 14.43&0.88 &1.12 & 1.24& KM27 &   PM \\
   419&  859.71&  813.91&  218.36& 16.29&0.11 &0.60& 0.72& KM131 &   PM \\
   449&  896.58&  866.32&  281.38& 16.44&0.12 &0.59 & 0.72& KM143 &   PM \\
   456&  730.79&  884.57&  221.72& 13.71&0.79 &1.07 & 1.14& KM17 &   PM \\
   465&  777.42&  897.74&  245.28& 16.07&0.45 &0.87 & 0.98& KM100 &   PM \\
   488&  625.73&  944.14&  288.85& 16.14&0.40 &0.83& 0.88& KM116 &   PM \\
   963&  699.83&  443.49&  221.51& 15.59&0.06 &0.57  & 0.69&KM58  &   PM \\
   974&  570.29&  464.89&  238.47& 13.70&0.77 &1.05 & 1.15& KM18 &   PM \\
   991&  824.76&  502.60&  204.79& 16.08&0.10 &0.60& 0.74& KM101 &   PM \\
  1061&  532.84&  609.89&  176.01& 16.27&0.97 &1.02 & 1.11& KM137 &   \\
  1077&  757.98&  637.12&   64.33& 16.24&0.10 &0.60 & 0.72&KM122  &   PM \\
  1089&  861.68&  652.43&  162.17& 16.16&0.27 &0.81  & 0.95& KM110 &   PM \\
  1101&  720.91&  668.72&   21.24& 16.15&0.12 &0.61& 0.71& KM112 &   PM \\
  1104&  707.22&  671.58&    9.77& 16.09&0.13 &0.61  & 0.73&KM105  &   PM \\
  1116&  751.69&  691.00&   57.86& 16.38&0.05 &0.64& 0.68& KM138 &   PM \\
  1128&  663.60&  710.11&   57.96& 16.04&0.19 &0.65 & 0.76& KM92 &   PM \\
  1132&  797.78&  710.48&  107.84& 16.35&0.11 &0.61  & 0.68& KM136 &   PM \\
  1147&  448.68&  737.15&  261.47& 16.21&0.14 &0.63& 0.77& KM128 &   PM \\
  1158&  649.41&  762.07&  109.46& 16.10&0.33 &0.71  & 0.82& KM117 &   PM \\
  1171&  878.02&  779.25&  211.53& 16.37&1.19 &1.46  & 1.99& KM140 &   \\
  1179&  821.92&  785.28&  171.27& 15.64&0.10 &0.59  & 0.70& KM57 &   PM \\
  1626&  682.58&  373.01&  292.51& 15.70&0.61 &1.00  & 1.09& KM61 &   PM \\
  1640&  828.25&  495.86&  212.26& 16.44&0.09 &0.59  & 0.77& KM149 &   PM \\
  1642&  875.89&  505.16&  237.67& 15.68&0.59 &0.98  & 1.09&KM62  &   PM \\
  1650&  753.47&  560.60&  117.30& 15.08&0.17 &0.65  & 0.70& KM39 &   \\
  1654&  925.16&  583.53&  239.45& 14.55&0.61 &0.99  & 1.15& KM29 &   PM \\
  1656&  725.49&  585.68&   83.32& 16.40&0.07 &0.61  & 0.71& KM146 &   PM \\
  1657&  662.72&  594.85&   79.44& 15.98&0.18 &0.73  & 0.83& KM88 & PM \\ \hline
\end{tabular} 
\newpage
\begin{tabular}{rrr rcc ccrc} \\ \\  \hline
Star &X & Y & Radius &$V$ &$(U-B)$ & $(B-V)$ & $(V-I)$ & Other &Membership \\
  &(pixel) &(pixel) &(pixel)&(mag)&(mag)&(mag)&(mag)&identification \\ \hline
  1661&  561.13&  604.88&  151.33& 16.06&0.42 &0.87  & 0.98& KM97 &   PM \\
  1662&  418.84&  605.15&  287.46& 15.24&1.58 &1.50  & 1.61& KM47 &   \\
  1663&  964.84&  611.62&  270.17& 16.10&0.24 &0.78  & 0.89& KM98 &   PM \\
  1666&  666.79&  627.02&   50.45& 15.95&0.07 &0.62  & 0.69& KM84 &   PM \\
  1668&  893.35&  645.43&  194.34& 15.35&0.33 &0.71  & 0.80& KM45 &   PM \\
  1675&  831.23&  665.10&  131.23& 16.22&0.06 &0.61  & 0.71& KM119 &   PM \\
  1685&  768.41&  695.85&   75.04& 13.64&0.72 &1.06  & 1.15&KM16  &   PM \\
  1691&  604.59&  718.30&  109.29& 15.15&0.32 &0.82  & 0.95& KM41 &   \\
  1701&  566.32&  766.05&  167.58& 16.17&0.10 &0.61  & 0.72& KM118 &   PM \\
  1702&  616.26&  765.31&  130.67& 15.80&0.62 &0.97  & 1.07& KM71 &   PM \\
  1714&  741.70&  830.99&  171.15& 15.82&0.12 &0.60  & 0.71& KM69 &   PM \\
  1717&  566.13&  838.96&  219.51& 15.96&0.88 &1.01 & 1.11&KM90  &   PM \\
  1718&  732.24&  853.39&  191.13& 16.27&0.14 &0.60  & 0.75& KM126 &   PM \\
  1895&  578.41&  396.63&  294.63& 15.46&0.58 &0.86  & 0.89& KM52 &   PM \\
  1937&  477.11&  518.17&  266.91& 16.43& &  & 0.71&  &   PM \\
  1948&  847.47&  538.41&  194.35& 13.42&0.83 &1.11 & 1.21&KM12  &   PM \\
  1986&  498.72&  632.23&  203.93& 16.35&0.34 &0.76  & 0.82&KM141  &   \\
  1996&  977.95&  643.70&  278.76& 16.15&0.53 &1.08  & 1.04&KM127  &   PM \\
  2009&  482.78&  667.41&  217.23& 16.44& &  & 0.72&  &   PM \\
  2034&  687.78&  743.96&   79.90& 16.38&0.19 &0.66  & 0.74& KM145 &   PM \\
  2239&  658.40&  647.67&   45.07& 15.36&0.65 &0.99  & 1.07& KM50 &   PM \\
  2247&  733.91&  736.48&   79.12& 14.51&0.46 &0.89  & 0.98& KM32 &   \\
  2260&  853.02&  906.84&  286.18& 16.14&0.28 &0.72  & 0.79& KM104 &   PM \\
  2307&  492.18&  455.51&  295.09& 15.14&0.35 &0.72  & 0.76& KM42 &   \\
 2313& 853.66&  539.06&  198.68& 16.14&0.10 &0.62  & 0.76& KM114 & PM \\ \hline 
\m{10}{c}{(B) Stars bluer than the MS turn-off (candidate for blue straggler)}\\ \hline
Star &X & Y & Radius &$V$ &$(U-B)$ & $(B-V)$ & $(V-I)$ & Other &Membership \\
  &(pixel) &(pixel) &(pixel)&(mag)&(mag)&(mag)&(mag)&identification \\ \hline
   169& 718.22& 376.70& 288.88&  15.86&0.17 &0.40 &0.48&KM72 & PM     \\ 
   292& 742.54& 615.41&  65.34&  15.97&0.16 &0.39 &0.41&KM82 & PM      \\
   303& 743.82& 631.21&  55.33&  16.07&0.07 &0.50 &0.60&KM102 &  PM      \\
   373& 684.03& 753.60&  90.03&  16.36&0.09 &0.54 &0.58&KM144 &  PM      \\
   413& 823.05& 806.89& 187.81&  16.00&0.11 &0.54 &0.63&KM85 &  PM      \\
   433& 468.83& 844.78& 292.85&  13.55&0.08 &0.54 &0.60&KM14 &      \\
   442& 638.28& 859.77& 204.32&  15.95&0.11 &0.51 &0.61&KM81 &  PM      \\
   446& 858.62& 863.88& 254.39&  14.44&0.12 &0.46 &0.52&KM26 &      \\
   507& 712.97& 963.58& 298.86&  13.23&0.18 &0.27 &0.25&KM9 &      \\
  1027& 545.34& 556.03& 189.19&  16.03&0.17 &0.49 &0.59&KM99 &  PM      \\
  1064& 849.30& 614.48& 157.62&  16.26&0.12 &0.49 &0.60&KM125 &  PM      \\
  1237& 564.34& 899.03& 270.51&  16.20&0.17 &0.47 &0.57&KM123 &  PM      \\
  1651& 616.62& 565.03& 130.18&  15.12&0.17 &0.22 &0.22&KM40 &      \\
  1680& 650.47& 675.41&  50.61&  16.26&0.08 &0.49 &0.57&KM129 &  PM      \\
  1684& 742.58& 694.04&  51.54&  15.81&0.00 &0.60 &0.61& KM68 &  PM      \\
  1700& 732.72& 758.40&  98.97&  14.57&0.16 &0.21 &0.15&KM30 &      \\
  1728& 728.99& 935.80& 272.35&  12.86&0.10 &0.60 &0.63&KM6 &      \\
  1911& 513.95& 458.03& 278.30&  13.39&0.07 &0.49 &0.55&KM13 &      \\
  1947& 733.43& 538.40& 130.94&  14.73&0.17 &0.29 &0.31&KM31 &      \\
  2041& 549.93& 781.97& 190.27&  16.29&0.14 &0.47 &0.54&KM131 &  PM      \\
  2062& 699.50& 848.78& 183.78&  15.57&0.16 &0.42 &0.43&KM55 &  PM      \\
  2079& 621.45& 898.18& 246.05&  15.36&0.11 &0.54 &0.62&KM48 &  PM      \\
  2400& 647.52& 621.46&  68.19&  14.13&0.09 &0.55 &0.62&KM23 &      \\ \hline
\end{tabular} 
\end{document}